\documentclass[twoside,a4paper,11pt,reqno]{amsart}
\usepackage{amsmath}%\usepackage{astron}%\usepackage[mathcal]{eucal}
\usepackage{tabularx}%\usepackage{rotating}
\numberwithin{equation}{section}

\newtheorem{theorem}{Theorem}[section]
\newtheorem{lemma}[theorem]{Lemma}
\theoremstyle{definition}

\theoremstyle{remark}

\numberwithin{equation}{section}

\begin{document}

\title[Global Solution to Enskog Equation]
{Global Solution to Enskog Equation with External Force in Infinite Vacuum}

\author{Zhenglu Jiang}
\address{Department of Mathematics, Zhongshan University,
Guangzhou 510275, P.~R.~China}
\email{mcsjzl@mail.sysu.edu.cn}
\thanks{This work  was supported by NSFC 10271121 and
the Scientific Research Foundation for the Returned Overseas Chinese
Scholars, the Education Ministry of China, and sponsored by joint
grants of NSFC 10511120278/10611120371 and RFBR 04-02-39026}

%\author{}
%\address{
%}
%\email{}
%\thanks{}

%    General info
\subjclass[2000]{76P05; 35Q75}

\date{\today.}


\keywords{The Enskog equation; global solution}

\begin{abstract}
We first give hypotheses of the bicharacteristic equations
corresponding to the Enskog equation with an external force. Since
the collision operator of the Enskog equation is more complicated
than that of the Boltzmann equation, these hypotheses are more
complicated than those given by Duan et {al.}~for the Boltzmann
equation. The hypotheses are very related to collision of particles
of moderately or highly dense gases along the bicharacteristic
curves and they can be used to make the estimation of the so-called
gain and loss integrals of the Enskog integral equation. Then, by
controlling these integrals, we show the existence and uniqueness of
the global mild solution to the Enskog equation in an infinite
vacuum for moderately or highly dense gases. Finally, we make some
remarks on the locally Lipschitz assumption
 of the collision factors in the Enskog equation.
\end{abstract}

\maketitle

\section{Introduction}
\label{intro} This paper is to consider the existence and uniqueness
of the global mild solution to the Enskog equation with an external
force in an infinite vacuum for moderately or highly dense gases.
Throughout this paper,
 ${\bf R}_+$ represents the positive side of the real axis including its origin and
 ${\bf R}^3$ denotes a three-dimensional Euclidean space.
In the presence of external forces depending on the time and space
variables, the Enskog equation is as follows (see \cite{cc70} or
\cite{e22}):
\begin{equation}
\frac {\partial f}{\partial t}+v\cdot\frac{\partial f}{\partial x}
+E(t,x)\cdot\frac{\partial f}{\partial v}=Q(f) \label{ee}
\end{equation}
where $ {f=f(t, x,v)}$ is a one-particle distribution function that
depends on the time $t\in {\bf R}_+,$ the position $x\in{\bf R}^3$
and the velocity $v\in{\bf R}^3,$ $E(t,x)$ is a vector-valued
function which belongs to ${\bf R}^3$ and represents an external
force with respect to the time and space variables,  and $Q$ is the
Enskog collision operator whose form will be explained below.

The collision operator $Q$ is expressed by the difference between
the gain and loss terms respectively defined by
 \begin{equation}
Q^+(f)(t,x,v)=\int_{{\bf R}^3\times
S_+^2}F^+(f)f(t,x,v^\prime)f(t,x-a\omega,w^\prime)B(v-w, \omega)
d\omega dw, \label{eekp}
\end{equation}
\begin{equation}
Q^-(f)(t,x,v)=\int_{{\bf R}^3\times
S_+^2}F^-(f)f(t,x,v)f(t,x+a\omega,w)B(v-w, \omega) d\omega dw.
\label{eekm}
\end{equation}
Here and below everywhere,  $S_+^2=\{\omega\in S^2 : \omega(v-w)\geq
0\}$ is a subset of a unit sphere surface $S^2$ in ${\bf R}^3,$ $a$
is a positive constant that represents a diameter of hard sphere,
$\omega$ is a unit vector along the line passing through the centers
of the spheres at their interaction, $B(v-w, \omega)=(v-w)\omega$ is
the collision kernel and
  $F^\pm$ are the collision factors.
  $F^\pm$ are usually assumed to be two functionals of $f,$
  more precisely speaking,
they depend on the density $\rho(t,x)=\int_{{\bf R}^3}f(t,x,v)dv$ at
the time $t$ and the point $x.$

In equations (\ref{eekp}) and (\ref{eekm}), $(v,w)$ and
$(v^\prime,w^\prime)$ are velocities before and after the collision,
respectively. As for the Boltzmann equation, the conservation of
both kinetic momentum and energy of two colliding particles gives
\begin{equation}
v+w= v^\prime +w^\prime,\hspace{1cm} v^2+w^2={v^\prime}^2
+{w^\prime}^2, \label{vrel01}
\end{equation}
which leads to their velocity relations
\begin{equation}
v^\prime=v-[(v-w)\omega]\omega,\hspace{1cm}w^\prime=w+[(v-w)\omega]\omega,
\label{vrel02}
\end{equation}
where $\omega\in S_+^2.$ We denote $u=v-w,$
$u_{\parallel}=(u\omega)\omega$  and $u_{\perp}=u-u_{\parallel}$
(see \cite{dyz05} or \cite{guo01}), thus getting another expression
of (\ref{vrel02}) as follows:
\begin{equation}
v^\prime=v-u_{\parallel},\hspace{1cm}w^\prime=v-u_{\perp}.
\label{vrel03}
\end{equation}
Then the gain and loss terms (\ref{eekp}) and (\ref{eekm}) can be
rechanged as
 \begin{equation}
Q^+(f)(t,x,v)=\int_{{\bf R}^3\times
S_+^2}F^+(f)f(t,x,v-u_{\parallel})f(t,x-a\omega,v-u_{\perp})B(u,
\omega) d\omega du, \label{eekp01}
\end{equation}
\begin{equation}
Q^-(f)(t,x,v)=\int_{{\bf R}^3\times
S_+^2}F^-(f)f(t,x,v)f(t,x+a\omega,v-u)B(u, \omega) d\omega du.
\label{eekm01}
\end{equation}

If the factors $F^\pm$ are set to be the same positive constant and
the diameter $a$ equal to zero in the density variables, then the
Enskog equation becomes the Boltzmann one that provides a successful
description for dilute gases. The Boltzmann equation
 is no longer valid for moderately or highly dense gases.
As a modification of the Boltzmann equation, the Enskog equation
proposed by Enskog \cite{e22} in 1922 is usually used to explain the
dynamical behavior of the density profile of moderately or highly
dense gases.

As we know, there are global solutions for the Boltzmann equation in
the absence of external forces not only in an infinite vacuum but
also with large initial data. Existence of such vacuum solutions was
first considered by Illner \& Shinbrot \cite{is84} and later by
Bellomo \& Toscani \cite{bt85}. The global existence of solution was
shown by DiPerna \& Lions \cite{dl} for the Boltzmann equation with
the large data. There are also
  some similar results about the Enskog equation without any effect of external
forces. For example, Polewczak \cite{p89} and Arkeryd \cite{a90}
gave their different existence proofs of global-in-time solutions to
the Enskog equation without external forces for both near-vacuum and
 large data, respectively. It is worth mentioning that
 some early works on the existence
of global solutions to the Enskog equation were given by Cercignani
 and/or Arkeryd (e.g. \cite{ac89}, \cite{ac90}, \cite{c87}, \cite{c88}).
 Recently, Duan et
{al.}~\cite{dyz05} proved the existence and uniqueness of a global
mild solutions for the Boltzmann equation with external forces in an
infinite vacuum and many relevant works can be found in the
reference. Now there is not yet such similar result for the Enskog
equation in the presence of external forces. The aim of this paper
is to extend the result to the case of the Enskog equation, that is,
to show the existence and uniqueness of such vacuum solution to the
Enskog equation (\ref{ee}) with (\ref{eekp}) and (\ref{eekm}). In
Section \ref{ao} hypotheses of the external forces are given and a
Banach space and its operators are constructed. These hypotheses are
very related to collision of particles of moderately or highly dense
gases along the bicharacteristic curves and they are much more
complicated than those given by Duan et {al.}~mentioned above for
the Boltzmann equation since the collision operator of the Enskog
equation is fairly more complicated than that of the Boltzmann
equation. In spite of this, the two examples shown by Duan et
{al.}~in \cite{dyz05} satisfy our hypotheses. Then the estimation of
the so-called gain and loss integrals is made in Section \ref{egli}.
An existence and uniqueness theorem of global mild solution to the
Enskog equation in an infinite vacuum is given in Section \ref{eu}
and some remarks on the assumption of the factors $F^\pm$ are
finally made in Section \ref{raf}.

\section{Hypotheses and Operators}
\label{ao} In this section we first give some constructive
hypotheses of the external forces with the help of the
bicharacteristic equations of the Enskog equation with these forces
and then build a Banach space and its operators relative to the
Enskog integral equation.

Let us begin with considering the bicharacteristic equations of the
Enskog equation (\ref{ee})
\begin{equation}
\frac{dX}{ds}=V,\hspace{1cm}\frac{dV}{ds}=E(s,X),\hspace{1cm}(X,V)|_{s=t}=(x,v).
\label{bceq}
\end{equation}
Suppose that such a vector-valued force function $E(t,x)$ allows
 the above system (\ref{bceq}) to have a global-in-time smooth solution denoted by
\begin{equation}
[X(s;t,x,v),V(s;t,x,v)] \label{bceqs}
\end{equation}
 for any fixed $(t,x,v)\in {\bf R}_+\times{\bf R}^3\times{\bf R}^3,$
and that there exist three functions $\alpha_i(s;t,x,v)$ ($i=1,2,3$)
such that the solution (\ref{bceqs}) satisfies the following
conditions:
\begin{equation}
\alpha_1(s;t,x,v)>0 ~\hbox{as}~s>0, ~\alpha_1(0;t,x,v)\geq 0,
~\alpha_2(s;t,x,v)>0, ~\alpha_3(s;t,x,v)\geq 0, \label{force01}
\end{equation}
\begin{equation}
X(0;s,X(s;t,x,v)+\xi,V(s;t,x,v)-\eta)=X(0;t,x,v)+\alpha_1(s;t,x,v)\eta+\alpha_2(s;t,x,v)\xi,
\label{force02}
\end{equation}
\begin{equation}
V(0;s,X(s;t,x,v)+\xi,V(s;t,x,v)-\eta)=V(0;t,x,v)-\alpha_2(s;t,x,v)\eta-\alpha_3(s;t,x,v)\xi,
\label{force03}
\end{equation}
\begin{equation}
\hbox{  either  } \alpha_3(s;t,x,v)\equiv 0  \hbox{  or  }
\max\{\alpha_1(s;t,x,v),\alpha_3(s;t,x,v)\}/\alpha_2(s;t,x,v)\leq\tau_0
, \label{force04}
\end{equation}
\begin{equation}
\min\{(\alpha_2(s;t,x,v))^2\alpha(s;t,x,v),\alpha_2(s;t,x,v)\alpha(s;t,x,v),\alpha_2(s;t,x,v)\}\geq
\alpha_0>0, \label{force05}
\end{equation}
for any $s\in {\bf R}_+$ and $(\xi,\eta)\in {\bf R}^3\times{\bf
R}^3$ when any point $(t,x,v)$ is fixed in $ {\bf R}_+\times{\bf
R}^3\times{\bf R}^3,$ where $\alpha_0,$ $e_0$ and $\tau_0$ are three
fixed positive
 constants independent of $s$ and $(t,x,v),$ and $\alpha(s;t,x,v)$ is denoted  by
\begin{equation}
\alpha(s;t,x,v)\equiv\alpha_1^\prime(s;t,x,v)\alpha_2(s;t,x,v)-\alpha_1(s;t,x,v)\alpha_2^\prime(s;t,x,v),
\label{alpha}
\end{equation}
 here
$\alpha_i^\prime(s;t,x,v)$ ($i=1,2$) represent the derivative with
respect to $s.$ For their understanding of their hypotheses of the
external force in the Boltzmann equation, Yuan et {al.}~\cite{dyz05}
took the following two examples: $E(t,x)=E_0(t)$ and
$E(t,x)=e_0^2x+E_0(t)$ with $e_0$ being a positive constant. Since
the Enskog equation is much more complicated than the Boltzmann one,
our hypotheses of the external forces for the Enskog equation are
much more complicated than those given by Yuan et {al.}~mentioned
above for the Boltzmann equation. In spite of this, the above two
examples satisfy the above hypotheses of the external force. These
examples are obviously suitable for our explanation of the
corresponding constructive conditions whether the Boltzmann equation
or the Enskog one is considered.

We give the five conditions (\ref{force01})-(\ref{force05}) in order
to get the following inequalities
\begin{equation}
|X(0;s,X(s;t,x,v),V(s;t,x,v)-u_{\parallel})|^2+
|X(0;s,X(s;t,x,v)-a\omega,V(s;t,x,v)-u_{\perp})|^2 \nonumber
\end{equation}
\begin{equation}
\geq
|X(0;t,x,v)|^2+|X(0;t,x,v)+\alpha_1(s;t,x,v)u-a\alpha_2(s;t,x,v)\omega|^2
\nonumber
\end{equation}
and
\begin{equation}
|V(0;s,X(s;t,x,v),V(s;t,x,v)-u_{\parallel})|^2+
|V(0;s,X(s;t,x,v)-a\omega,V(s;t,x,v)-u_{\perp})|^2 \nonumber
\end{equation}
\begin{equation}
\geq
|V(0;t,x,v)|^2+|V(0;t,x,v)-\alpha_2(s;t,x,v)u+a\alpha_3(s;t,x,v)\omega|^2
\nonumber
\end{equation}
along their bicharacteristic curves after collision of particles of
moderately or highly dense gases. The above two inequalities can be
used to control the gain and loss integral terms of the Enskog
integral equation along the bicharacteristic curves. Therefore this
form of the external forces is pertinent to collision of particles
of moderately or highly dense gases along the bicharacteristic
curves so that the gain and loss integral terms can be estimated.

We below give a representation of mild solution to the Enskog
equation. Let us first introduce a notation $f^\#$  defined as
$$ f^\#(s;t, x,v)=f(s,X(s;t,x,v),V(s;t,x,v) )$$ for any measurable function $f$ on
${\bf R}_+{\times}{\bf R}^3{\times}{\bf R}^3.$ Obviously, it can be
found that $ f^\#(t;t, x,v)=f(t,x,v)$ and that $ f^\#(0;t,
x,v)=f(0,X(0;t,x,v),V(0;t,x,v)).$

Along the bicharacteristic curves, the Enskog equation (\ref{ee})
can be also written as
$$\frac{d}{ds}f^\#(s;t, x,v)=Q(f)^\#(s;t, x,v),$$
which leads to the following integral equation
\begin{equation}
f(t, x,v)=f_0(X(0;t,x,v),V(0;t,x,v))+\int_0^tQ(f)^\#(s;t, x,v)ds,
\label{mild}
\end{equation}
where $f_0(x,v)\equiv f(0,x,v).$ A function $f(t, x, v)$ is called
global mild solution to the Enskog equation (\ref{ee}) if $f(t, x,
v)$ satisfies the above integral equation (\ref{mild}) for almost
every $(t, x, v)\in {\bf R}_+\times{\bf R}^3\times{\bf R}^3.$

Then  we construct a subset $M$ of a Banach space $C({\bf
R}_+\times{\bf R}^3\times{\bf R}^3),$ which has the property that
every element $f=f(s,x,v)\in M$ if and only if there exists a
positive constant $c$ such that $f$ satisfies
$$|f(t,x,v)|\leq c h(X(0;t,x,v))m(V(0;t,x,v))$$
where
\begin{equation}
h(x)=e^{-p|x|^2},~~~m(v)=e^{-qv^2}, \label{a01ex}
\end{equation}
for any fixed $p$ and $q$ in $(0,+\infty).$ It follows that $M$ is a
Banach space when it has a norm of the following form
$$||f||=\sup\limits_{t,x,v}\{|f(t,x,v)|h^{-1}(X(0;t,x,v))m^{-1}(V(0;t,x,v))\}.$$
In particular,
$||f(0,x,v)||=\sup\limits_{x,v}\{|f(0,x,v)|h^{-1}(x)m^{-1}(v)\}.$
The initial data $f_0\equiv f(0,x,v)$ is bounded in $L^1({\bf
R}^3\times{\bf R}^3).$ This implies that the total mass of the
system is finite. Hence the mean free path is sufficiently large if
the finite total mass is sufficiently small. This is exactly the
requirement on the Enskog equation with external forces in an
infinite vacuum, which is similar to one considered by Illner \&
Shinbrot \cite{is84} for the Boltzmann equation. It is worth
mentioning that in other cases there are many different classes of
functions which can be taken as the choice of $h(x)$ and $ m(v)$
(see \cite{dyz05}, \cite{guo01}). For example, in the case of the
external forces depending only on the time $t,$ one can also choose
$h(x)=(1+|x|^2)^{-p}$ and $ m(v)=e^{-qv^2}$ for any fixed $p\in
(1/2,+\infty)$ and $q\in (0,+\infty);$ when $p>3/2,$ the initial
data $f_0\equiv f(0,x,v)$ is bounded in $L^1({\bf R}^3\times{\bf
R}^3);$ when $1/2<p\leq 3/2,$ the initial total mass might be
infinite; this choice of  both $h(x)$ and $ m(v)$ is not suitable
for using our method considered in this paper to deal with this
existence problem of vacuum solutions to the Enskog equation with
external forces depending on the time and space variables, however,
this choice can be applied to the case of the Boltzmann equation in
the presence of external forces \cite{dyz05}.

To give global existence, it is necessary to study the properties of
the collision operator in a Banach space.  To do this, by
(\ref{eekp01}) and (\ref{eekm01}), $Q(f)^\#(s;t, x, v)$ can be first
rewritten as the difference between the gain and loss terms of other
two forms
 \begin{eqnarray}
Q^+(f)^\#(s;t,x,v)  =
\int_{{\bf R}^3\times S_+^2}F^+(f)f(s,X(s;t,x,v),V(s;t,x,v)-u_{\parallel}) \nonumber \\
\times f(s,X(s;t,x,v)-a\omega,V(s;t,x,v)-u_{\perp})B(u, \omega)
d\omega du, \label{eekph}
\end{eqnarray}
\begin{eqnarray}
Q^-(f)^\#(s;t,x,v) =\int_{{\bf R}^3\times S_+^2}F^-(f)f(s,X(s;t,x,v),V(s;t,x,v))\nonumber \\
\times f(s,X(s;t,x,v)+a\omega,V(s;t,x,v)-u)B(u, \omega) d\omega du.
\label{eekmh}
\end{eqnarray}

Estimation of the collision integrals can be then made by use of
 a similar argument to that developed in the previous work
(see \cite{dyz05}, \cite{p89},  \cite{t86}).  According to
(\ref{eekph}) and (\ref{eekmh}), we in fact have to estimate the
following two integrals:
\begin{equation}
I_g\equiv\int_0^t\int_{{\bf R}^3\times S_+^2}
h(X(0;s,X(s;t,x,v),V(s;t,x,v)-u_{\parallel}))
 \nonumber\end{equation}
\begin{equation}\times m(V(0;s,X(s;t,x,v),V(s;t,x,v)-u_{\parallel}))
h(X(0;s,X(s;t,x,v)-a\omega,V(s;t,x,v)-u_{\perp}))
 \nonumber\end{equation}
\begin{eqnarray}
\times
m(V(0;s,X(s;t,x,v)-a\omega,V(s;t,x,v)-u_{\perp}))B(u,\omega)d\omega
duds, \label{gain}
\end{eqnarray}
\begin{equation}
I_l\equiv\int_0^t\int_{{\bf R}^3\times S_+^2}
h(X(0;s,X(s;t,x,v),V(s;t,x,v))) \nonumber\end{equation}
\begin{equation}\times m(V(0;s,X(s;t,x,v),V(s;t,x,v)))h(X(0;s,X(s;t,x,v)+a\omega,V(s;t,x,v)-u))
\nonumber\end{equation}
\begin{eqnarray}
\times m(V(0;s,X(s;t,x,v)+a\omega,V(s;t,x,v)-u))B(u,\omega)d\omega
duds. \label{loss}
\end{eqnarray}
Here $I_g$ and $I_l$ are called the gain and loss integrals
respectively. Once the estimation of the integrals is finished, the
global existence result may be obtained by constructing a
contractive map from a Banach space to itself. Therefore it is one
of the best important to estimate the above two integrals,
especially  the gain one. It will be discussed in the next section.

We finally denote an operator $J$ on $M$ by
\begin{equation}
J(f)=f_0(X(0;t,x,v), V(0;t,x,v))+\int_0^tQ(f)^\#(s;t, x, v)ds.
\label{op}
\end{equation}
It will be proved in Section \ref{eu} that $J$ is indeed a
contractive map on a Banach space. This is what we need.

\section{Estimation of the Gain and Loss Integrals}
\label{egli} In this section the so-called gain and loss integrals
are estimated by use of a similar device to one given in \cite{p89}
for the Enskog equation in the absence of external forces. To do
this, we first introduce the preliminary lemmas which will be used
below.
\begin{lemma}
Let any $z\in {\bf R}^3,$ $s\in {\bf R}_+,$
$(u_{\parallel},u_{\perp})\in {\bf R}^3\times{\bf R}^3$ with
$u_{\parallel}u_{\perp}=0$  and $\omega\in S^2$ with
$u_{\parallel}\omega\geq 0.$ Then
\begin{equation}
|z\pm su_{\parallel}|^2+|z\pm su_{\perp}\mp
a\omega|^2\geq|z|^2+|z\pm s(u_{\parallel}+u_{\perp})\mp a\omega|^2
\label{lem01eq}\end{equation} for any fixed real number $a\in {\bf
R}_+.$ \label{lem01}
\end{lemma}
\begin{lemma}
Let $p>0$  and $(z,u)\in {\bf R}\times{\bf R}$ with $u\not=0.$ Then
\begin{equation}
\int_0^{+\infty}e^{-p|z+su|^2}ds\leq \frac{\sqrt{\pi}}{\sqrt{p}|u|}.
\label{lem02eq}\end{equation} \label{lem02}
\end{lemma}
\begin{lemma}
Let $q>0,$  $-2<\gamma\leq1$ and $z\in {\bf R}^3 .$ Then
\begin{equation}
\int_{{\bf R}^3}|u|^{\gamma-1}e^{-q|z-u|^2}du\leq
\frac{4\pi}{\gamma+2}+\frac{\pi}{q^{3/2}}.
\label{lem03eq}\end{equation} \label{lem03}
\end{lemma}
\begin{lemma}
Let  any  $(s,z,u)\in {\bf R}\times{\bf R}^3\times{\bf R}^3$ and
$\omega_{\perp}$ be a unit vector perpendicular to $\omega \in S^2.$
Then
\begin{equation}
|z+su+h\omega|\geq |z\omega_{\perp}+su\omega_{\perp}|
\label{lem04eq}\end{equation} for any $h\in {\bf R}.$ \label{lem04}
\end{lemma}

The proof of Lemmas \ref{lem01} and \ref{lem04} is easily given.
Lemmas \ref{lem02} and \ref{lem03} can be obtained from the
transformation of integral variables.

\begin{lemma}
Assume that three functions $\alpha_i(s;t,x,v)$ ($i=1,2,3$) satisfy
 the external force conditions (\ref{force01}) and (\ref{force05}), and that
$$\tilde{I}_l(z_1,z_2,t,x,v)\equiv
\int_0^t\int_{{\bf R}^3\times
S_+^2}|uw|e^{-p|z_1+\alpha_1(s;t,x,v)u+a\alpha_2(s;t,x,v)\omega|^2}
$$$$\times e^{-q|z_2-\alpha_2(s;t,x,v)u-a\alpha_3(s;t,x,v)\omega|^2}dud\omega ds$$
for any fixed real number $a\in {\bf R}_+.$ Then
\begin{equation}
\tilde{I}_l(z_1,z_2,t,x,v)\leq \tilde{I}_{lpq} \label{lem05eq}
\end{equation}
 for any $(z_1,z_2)\in {\bf R}^3\times{\bf R}^3,$ $(t,x,v)\in{\bf R}_+\times{\bf R}^3\times{\bf R}^3,$
where $\tilde{I}_{lpq}$ is a positive constant depending only on $p$
and $q,$ $p>0,$ $q>0.$ \label{lem05}
\end{lemma}
\begin{proof}
First let us fix $(z_1,z_2)\in {\bf R}^3\times{\bf R}^3$ and
$(t,x,v)\in{\bf R}_+\times{\bf R}^3\times{\bf R}^3.$
 Put
$\bar{u}=\alpha_2(s;t,x,v)u+a\alpha_3(s;t,x,v)\omega.$   Then
$\bar{u}\omega=\alpha_2(s;t,x,v)(u\omega)\omega+a\alpha_3(s;t,x,v)$
for $\omega \in S_+^2.$ By (\ref{force01}), it thus follows that
\begin{eqnarray}
\tilde{I}_l(z_1,z_2,t,x,v)\leq \int_0^t\int_{{\bf R}^3\times
S_+^2}|\bar{u}\omega|(\alpha_2(s;t,x,v))^{-4}
\hspace*{0.2\linewidth}\nonumber\\
 \times e^{-p|z_1+\frac{\alpha_1(s;t,x,v)}{\alpha_2(s;t,x,v)}\bar{u}+\frac{a\widetilde{\alpha}(s;t,x,v)}{\alpha_2(s;t,x,v)}
\omega|^2}e^{-q|z_2-\bar{u}|^2}d\bar{u}d\omega ds, \label{lem05peq}
\end{eqnarray}
where
$\widetilde{\alpha}(s;t,x,v)=\alpha_2^2(s;t,x,v)-\alpha_1(s;t,x,v)\alpha_3(s;t,x,v).$
Take $\tau=\frac{\alpha_1(s;t,x,v)}{\alpha_2(s;t,x,v)}.$  Then, by
(\ref{force05}),
$\frac{d\tau}{ds}=\frac{\alpha(s;t,x,v)}{(\alpha_2(s;t,x,v))^2}>0.$
By replacing the integral variable $s$ with the new variable $\tau$
and using Lemma \ref{lem04}, the estimation of the integral on the
right side of (\ref{lem05peq}) thus gives
\begin{equation}
\tilde{I}_l(z_1,z_2,t,x,v)\leq \int_0^{+\infty}\int_{{\bf R}^3\times
S_+^2}|\bar{u}\omega|(\alpha_2(s;t,x,v))^{-2}(\alpha(s;t,x,v))^{-1}
\nonumber
\end{equation}
\begin{equation}
\times e^{-p|z_1\omega_{\perp}+\tau\bar{u}\omega_{\perp}|^2}
e^{-q|z_2-\bar{u}|^2}d\bar{u}d\omega d\tau \nonumber \\
\end{equation}
\begin{equation}
\leq\frac{1}{\alpha_0} \int_{{\bf R}^3\times
S_+^2}|\bar{u}\omega|\left\{\int_0^{+\infty}
e^{-p|z_1\omega_{\perp}+\tau\bar{u}\omega_{\perp}|^2}d\tau\right\}
e^{-q|z_2-\bar{u}|^2}d\bar{u}d\omega, \label{lem05pineq}
\end{equation}
where $ \omega_{\perp}$ is a unit vector perpendicular to $\omega.$
The last inequality in (\ref{lem05pineq}) comes from (\ref{force05})
and Fubini's theorem. By estimation of the integral on the right
side of the last inequality, it then follows that
  \begin{equation}
\tilde{I}_l(z_1,z_2,t,x,v)\leq\frac{1}{\alpha_0}\frac{\sqrt{\pi}}{\sqrt{p}}
\int_{{\bf R}^3\times S_+^2}
\frac{|\bar{u}\omega|}{|\bar{u}\omega_{\perp}|}
e^{-q|z_2-\bar{u}|^2}d\bar{u}d\omega \leq
\frac{1}{\alpha_0}\frac{4\pi^{5/2}
}{\sqrt{p}}\left(\frac{4}{3}+\frac{1}{q^{3/2}}\right),
\label{lem05pineq01}
\end{equation}
where Lemmas \ref{lem02} and \ref{lem03} are used. The proof of
Lemma \ref{lem05} is hence completed.
\end{proof}
\begin{lemma}
Assume that three functions $\alpha_i(s;t,x,v)$ ($i=1,2,3$) satisfy
 the external force conditions (\ref{force01}), (\ref{force04}) and (\ref{force05}), and put
$$\tilde{I}_g(z_1,z_2,t,x,v)\equiv
\int_0^t\int_{{\bf R}^3\times
S_+^2}|u\omega|e^{-p|z_1+\alpha_1(s;t,x,v)u-a\alpha_2(s;t,x,v)\omega|^2}
$$$$
\times e^{-q|z_2-\alpha_2(s;t,x,v)u+a\alpha_3(s;t,x,v)\omega|^2}
dud\omega ds,$$ where $a\geq 0.$ Then
\begin{equation}
\tilde{I}_g(z_1,z_2,t,x,v)\leq \tilde{I}_{gpq} \label{lem06eq}
\end{equation}
 for any $(z_1,z_2)\in {\bf R}^3\times{\bf R}^3,$ $(t,x,v)\in{\bf R}_+\times{\bf R}^3\times{\bf R}^3,$
where $\tilde{I}_{gpq}$ is a positive constant depending only on $p$
and $q,$ $p>0,$ $q>0.$ \label{lem06}
\end{lemma}
\begin{proof}
Let us fix $(z_1,z_2)\in {\bf R}^3\times{\bf R}^3$ and
$(t,x,v)\in{\bf R}_+\times{\bf R}^3\times{\bf R}^3.$ Note that
$\alpha_2(s;t,x,v)>0$ in (\ref{force01}). Put
$\bar{u}=\bar{u}_{\parallel}+\bar{u}_{\perp},$ where
$\bar{u}_{\parallel}=\alpha_2(s;t,x,v)u_{\parallel}-a\alpha_3(s;t,x,v)\omega$
and $\bar{u}_{\perp}=\alpha_2(s;t,x,v)u_{\perp}.$ Then
$\bar{u}\omega=\alpha_2(s;t,x,v)(u\omega)-a\alpha_3(s;t,x,v)$  for
$\omega \in S_+^2.$
 Thus
\begin{equation}
\tilde{I}_g(z_1,z_2,t,x,v)\leq \int_0^t\int_{{\bf R}^3\times
S_+^2}(|\bar{u}\omega|+a\alpha_3(s;t,x,v))(\alpha_2(s;t,x,v))^{-4}
\nonumber
\end{equation}
\begin{equation}
\times e^{-p|z_1+
\frac{\alpha_1(s;t,x,v)}{\alpha_2(s;t,x,v)}\bar{u}-\frac{a\widetilde{\alpha}(s;t,x,v)}{\alpha_2(s;t,x,v)}
\omega|^2} e^{-q|z_2-\bar{u}|^2}d\bar{u}d\omega ds, \nonumber
\end{equation}
where
$\widetilde{\alpha}(s;t,x,v)=\alpha_2^2(s;t,x,v))-\alpha_1(s;t,x,v)\alpha_3(s;t,x,v).$
To prove this lemma,  it suffices to consider the case of
$\max\{\alpha_1(s;t,x,v),\alpha_3(s;t,x,v)\}/\alpha_2(s;t,x,v)\leq\tau_0$
in (\ref{force01}). By repeating a similar integral estimation to
one given in Lemma \ref{lem05}, the estimate of the integrals on the
right side of the above inequality gives
\begin{equation}
\tilde{I}_g(z_1,z_2,t,x,v)
 \stackrel{\hbox{(a)}}{\leq}
 \int_0^{\tau_0}\int_{{\bf R}^3\times S_+^2}(|\bar{u}\omega|+a\alpha_3(s;t,x,v))
(\alpha_2(s;t,x,v))^{-2}(\alpha(s;t,x,v))^{-1} \nonumber
\end{equation}
\begin{equation}
\times e^{-p|z_1\omega_{\perp}+\tau\bar{u}\omega_{\perp}|^2}
e^{-q|z_2-\bar{u}|^2}d\bar{u}d\omega d\tau \nonumber
\end{equation}
\begin{equation}
 \stackrel{\hbox{(b)}}{\leq}
\frac{1}{\alpha_0} \int_{{\bf R}^3\times
S_+^2}\left\{|\bar{u}\omega|\int_0^{+\infty}e^{-p|z_1\omega_{\perp}+\tau\bar{u}\omega_{\perp}|^2}d\tau+a\tau_0^2\right\}
e^{-q|z_2-\bar{u}|^2}d\bar{u}d\omega \nonumber
\end{equation}
\begin{equation}
 \stackrel{\hbox{(c)}}{\leq} \frac{1}{\alpha_0}
\int_{{\bf R}^3\times
S_+^2}\left\{\frac{\sqrt{\pi}}{\sqrt{p}}\frac{|\bar{u}\omega|}{|\bar{u}\omega_{\perp}|}
+a\tau_0^2\right\}e^{-q|z_2-\bar{u}|^2}d\bar{u}d\omega
 \stackrel{\hbox{(d)}}{\leq}  \frac{4\pi^2}{\alpha_0}\left(\frac{\sqrt{\pi}}{\sqrt{p}}+a\tau_0^2\right)(\frac{4}{3}+\frac{1}{q^{3/2}}),
 \label{lem06peq}
\end{equation}
for $\alpha_3(s;t,x,v)\not= 0,$ where $\omega_{\perp}$ is a unit
vector perpendicular to $\omega,$ (a) is obtained by first making
the transformation
$\tau=\frac{\alpha_1(s;t,x,v)}{\alpha_2(s;t,x,v)}$ and then using
(\ref{force04}) and  Lemma \ref{lem04}, (b) is given by
(\ref{force05}),  (c) is obtained by Lemma \ref{lem02} and (d)
results from Lemma \ref{lem03}. This hence completes the proof of
Lemma \ref{lem06}.
\end{proof}
By Lemmas \ref{lem05} and \ref{lem06}, we can hence give the
estimates of the gain and loss integrals as follows.
\begin{lemma}
Let $I_g$ and $I_l$ be the same integrals as defined by (\ref{gain})
and (\ref{loss}), respectively. In the integrals, $a$ is a positive
constant. Suppose that all the five external force conditions
(\ref{force01})--(\ref{force05}) hold for these functions
$X(s;t,x,v)$ and $V(s;t,x,v)$ defined by the solution (\ref{bceqs})
to the system (\ref{bceq}), and that $h(x)$ and $m(v)$ are the same
as in (\ref{a01ex}). Then it follows that
\begin{equation}
I_g\leq Kh(X(0;t,x,v))m(V(0;t,x,v)), \label{a01}
\end{equation}
\begin{equation}
I_l \leq Kh(X(0;t,x,v))m(V(0;t,x,v)), \label{a02}
\end{equation}
for any $(t,x,v)\in{\bf R}_+\times{\bf R}^3\times{\bf R}^3$ and some
positive constant $K.$ \label{estimate}
\end{lemma}
\begin{proof}
Let us first estimate the loss integral. By using (\ref{force02})
and (\ref{force03}), the loss integral (\ref{loss}) can be rewritten
as
\begin{equation}
I_l=\int_0^t\int_{{\bf R}^3\times
S_+^2}|u\omega|e^{-p|X(0;t,x,v)|^2}
e^{-p|X(0;t,x,v)+\alpha_1(s;t,x,v)u+a\alpha_2(s;t,x,v)\omega|^2}\nonumber
\end{equation}
\begin{equation}
\times e^{-q|V(0;t,x,v)|^2}
e^{-q|V(0;t,x,v)-\alpha_2(s;t,x,v)u-a\alpha_3(s;t,x,v)\omega|^2}
dud\omega ds \nonumber
\end{equation}
\begin{equation}
=
h(X(0;t,x,v))m(V(0;t,x,v))\widetilde{I}_l(X(0;t,x,v),V(0;t,x,v),t,x,v)).
\label{loss01}
\end{equation}
It follows from Lemma \ref{lem05} that
\begin{equation}
I_l\leq h(X(0;t,x,v))m(V(0;t,x,v))\widetilde{I}_{lpq}.
\label{loss02}
\end{equation}

Then the estimation of the gain integral will be made below.
Similarly, by using  (\ref{force02}) and (\ref{force03}), the gain
integral (\ref{gain}) is as follows:
\begin{equation}
I_g=\int_0^t\int_{{\bf R}^3\times
S_+^2}|u\omega|e^{-p|X(0;t,x,v)+\alpha_1(s;t,x,v)u_{\parallel}|^2}
 e^{-p|X(0;t,x,v)+\alpha_1(s;t,x,v)u_{\perp}-a\alpha_2(s;t,x,v)\omega|^2}
\nonumber
\end{equation}
\begin{equation}
\times e^{-q|V(0;t,x,v)-\alpha_2(s;t,x,v)u_{\parallel}|^2}
 e^{-q|V(0;t,x,v)-\alpha_2(s;t,x,v)u_{\perp}+a\alpha_3(s;t,x,v)\omega|^2}
dud\omega ds. \label{gain01}
\end{equation}
Using Lemma \ref{lem01}, we have
\begin{equation}
I_g\leq e^{-p|X(0;t,x,v)|^2}e^{-q|V(0;t,x,v)|^2}\int_0^t\int_{{\bf
R}^3\times S_+^2}|u\omega|
e^{-p|X(0;t,x,v)+\alpha_1(s;t,x,v)u-a\alpha_2(s;t,x,v)\omega|^2}\nonumber
\end{equation}
\begin{equation}
\times
e^{-q|V(0;t,x,v)-\alpha_2(s;t,x,v)u+a\alpha_3(s;t,x,v)\omega|^2}
dud\omega ds\nonumber
\end{equation}
\begin{equation}
=h(X(0;t,x,v))m(V(0;t,x,v))\widetilde{I}_g(X(0;t,x,v),V(0;t,x,v),t,x,v).
\label{gain02}
\end{equation}
It follows from Lemma \ref{lem06} that
\begin{equation}
I_g\leq h(X(0;t,x,v))m(V(0;t,x,v))\widetilde{I}_{gpq}.
\label{gain03}
\end{equation}

The proof of Lemma \ref{estimate} is hence completed.
\end{proof}

\section{Existence and Uniqueness}
\label{eu} In this section we show a result about the existence and
uniqueness of such vacuum solution to the Enskog equation in
presence of external forces. To do this, we first define a set $M_R$
by
\begin{equation}
M_R=\{f:||f||\leq R, f\in C({\bf R}_+\times {\bf R}^3\times {\bf
R}^3 \} \label{subset}
\end{equation}
and then assume that $F^\pm$ are two functionals on $M_R$ such that
the so-called locally Lipschitz condition
\begin{equation}
|F^\pm(f)-F^\pm(g)|\leq L(R)||f-g|| \label{factorf}
\end{equation}
holds for any $f,g\in M_R$ where $M_R$ is defined by (\ref{subset})
and $L(R)$ is a positive nondecreasing function on ${\bf R}_+.$ Thus
we can get the following lemma.
\begin{lemma}
Suppose that all the five conditions
(\ref{force01})--(\ref{force05}) hold for any $X(s;t,x,v)$ and
$V(s;t,x,v)$ defined by the solution (\ref{bceqs}) to the system
(\ref{bceq}), and that the factors $F^\pm$ in the collision
integrals $Q^\pm(f)^\#(s;t,x,v)$ defined by (\ref{eekph}) and
(\ref{eekmh}) are two functionals satisfying the inequality
(\ref{factorf}). Let $h(x)$ and $m(v)$ be the same as in
(\ref{a01ex}).  In the collision integrals, $a$ is a positive
constant. Then the following inequalities hold:
$$\int_0^t|Q^+(f)^\#(s;t,x,v)|ds\leq C(R)h(X(0;t,x,v))m(V(0;t,x,v))||f||^2,$$
$$\int_0^t|Q^-(f)^\#(s;t,x,v)|ds\leq C(R)h(X(0;t,x,v))m(V(0;t,x,v))||f||^2$$
for any $f\in M_R,$ where $C(R)$ is a positive nondecreasing
function on ${\bf R}_+.$ \label{lem}
\end{lemma}
\begin{proof}
It can be first found from the assumption (\ref{factorf}) of the two
functionals $F^\pm$ that there exists a positive constant
$\tilde{L}(R)=L(R)R+|F^+(0)|+|F^-(0)|$ such that $|F^\pm(f)| \leq
\tilde{L}(R)$ for any $f\in M_R.$ It follows from  (\ref{eekph}) and
(\ref{eekmh}) that
 \begin{equation}
\int_0^tQ^+(f)^\#(s;t,x,v)ds \leq\tilde{L}(R)\int_0^t\int_{{\bf
R}^3\times S_+^2}
 ||f||^2h(X(0;s,X(s;t,x,v),V(s;t,x,v)-u_{\parallel}))
 \nonumber
\end{equation}
\begin{equation}
\times
m(V(0;s,X(s;t,x,v),V(s;t,x,v)-u_{\parallel}))h(X(0;s,X(s;t,x,v)-a\omega,V(s;t,x,v)-u_{\perp}))
\nonumber
\end{equation}
\begin{equation}
\times m(V(0;s,X(s;t,x,v)-a\omega,V(s;t,x,v)-u_{\perp})) u\omega
d\omega duds, \label{rbeekphintt}
\end{equation}
\begin{equation}
\int_0^tQ^-(f)^\#(s;t,x,v)ds \leq\tilde{L}(R)\int_0^t\int_{{\bf
R}^3\times S_+^2}
 ||f||^2h(X(0;s,X(s;t,x,v),V(s;t,x,v)))\nonumber
\end{equation}
\begin{equation}
\times m(V(0;s,X(s;t,x,v),V(s;t,x,v)))
h(X(0;s,X(s;t,x,v)+a\omega,V(s;t,x,v)-u)) \nonumber
\end{equation}
\begin{equation}
\times m(V(0;s,X(s;t,x,v)+a\omega,V(s;t,x,v)-u)) u\omega d\omega
duds. \label{rbeekmhintt}
\end{equation}
By (\ref{a01}) and (\ref{a02}), (\ref{rbeekphintt}) and
(\ref{rbeekmhintt}) give
\begin{eqnarray}
\int_0^tQ^+(f)^\#(\tau,x,v) d\tau \leq
\tilde{L}(R)Kh(X(0;t,x,v))m(V(0;t,x,v))||f||^2, \nonumber
\end{eqnarray}
\begin{eqnarray}
\int_0^tQ^-(f)^\#(\tau,x,v)d\tau \leq
\tilde{L}(R)Kh(X(0;t,x,v))m(V(0;t,x,v))||f||^2. \nonumber
\end{eqnarray}
Take $C(R)=\tilde{L}(R)K.$  It obviously follows that Lemma
\ref{lem} holds.
\end{proof}

Then we can get the following theorem.
\begin{theorem}
Suppose that all the five external force conditions
(\ref{force01})--(\ref{force05}) hold for these functions
$X(s;t,x,v)$ and $V(s;t,x,v)$ defined by the solution (\ref{bceqs})
to the system (\ref{bceq}), and that the factors $F^\pm$ in the
collision integrals $Q^\pm(f)(t,x,v)$ defined by (\ref{eekp}) and
(\ref{eekm}) are two functionals satisfying the inequality
(\ref{factorf}). In the collision integrals, $a$ is a positive
constant.
 Then there exists a positive constant
$R_0$ such that the  Enskog equation (\ref{ee}) with (\ref{eekp})
and (\ref{eekm})
 has a unique  non-negative global mild solution $f=f(t,x,v)\in M_{R_0}$
 through a non-negative initial data $f_0=f_0(x,v)$ when
$$\sup\limits_{t,x,v}\{f_0(X(0;t,x,v),V(0;t,x,v))h^{-1}(X(0;t,x,v))m^{-1}(V(0;t,x,v))\} $$ is
 sufficiently small,  where $h(x)$ and $m(v)$ are the same as in (\ref{a01ex}).
\label{th}
\end{theorem}
Theorem \ref{th} shows that there exists a unique  global mild
solution to the Enskog equation (\ref{ee}) given by (\ref{eekp}) and
(\ref{eekm}) with the initial data near vacuum if a suitable
assumption of the external force is given. As in \cite{j07}, we
below give our proof of Theorem \ref{th}.

\begin{proof}
By (\ref{op}) and Lemma \ref{lem}, we have
\begin{equation}
\begin{array}{c}
|J(f)|h^{-1}(X(0;t,x,v))m^{-1}(V(0;t,x,v))  \\
\leq |f_0(X(0;t,x,v),V(0;t,x,v))
|h^{-1}(X(0;t,x,v))m^{-1}(V(0;t,x,v))+2C(R)||f||^2  \\
\leq R/2+2C(R)R^2
\end{array}
\nonumber
\end{equation}
for any $f\in M_R$ and $f_0$ with $||f_0||\leq R/2.$ Since $C(R)$ is
a positive nondecreasing function on ${\bf R}_+,$ it follows that
$||J(f)||\leq R$ for sufficiently small $R.$ Therefore $J$ is an
operator from $M_R$ to itself for sufficiently small $R.$ Similarly,
it can be also found that $J$ is a contractive operator on $M_R$ for
some suitably small $R.$ Thus there exists a unique element $f\in
M_R$ such that $f=J(f),$ i.e., (\ref{mild}) holds. It then follows
from the same argument as the one in \cite{is84} (or see
\cite{ks78}, \cite{ukai86}) that if $f_0(x, v)\geq 0$ then $f(t,x,
v)\geq 0.$ Hence the proof of Theorem \ref{th} is finished.
\end{proof}

\section{Remarks on the Assumption of the Factors $F^\pm$}
\label{raf} In this section we make some remarks on the locally
Lipschitz assumption (\ref{factorf})
 of the factors of $F^\pm$ appearing in Theorem \ref{th}
given in the previous section.

We begin with a different kind of  locally Lipschitz condition of
$F^\pm.$ It was originally given by Polewczak \cite{p89} as follows:
\begin{equation}
|F^\pm(f)-F^\pm(g)|\leq L_0(R)\left|\int_{{\bf
R}^3}f(t,x,v)dv-\int_{{\bf R}^3}g(t,x,v)dv\right | \label{factorf01}
\end{equation}
holds for any $f=f(t,x,v),g=g(t,x,v)\in M_R$ where $M_R$ is defined
by (\ref{subset}) and $L_0(R)$ is a positive nondecreasing function
on ${\bf R}_+.$ Note that assumptions (\ref{force02}) and
(\ref{force03}) have the following properties:
\begin{equation}
X(0;t,x+\xi,v)-X(0;t,x,v)=\alpha_2(s;t,x,v)\xi \label{force02s}
\end{equation}
and
\begin{equation}
V(0;t,x,v+\eta)-V(0;t,x,v)=\alpha_2(s;t,x,v)\eta \label{force03s}
\end{equation}
for any $(\xi,\eta)\in {\bf R}^3\times{\bf R}^3$ when any point
$(t,x,v)$ is fixed in ${\bf R}_+\times{\bf R}^3\times{\bf R}^3.$ By
(\ref{force02s}) and (\ref{force03s}), we have
\begin{equation}
\frac{\partial X(0;t,x,v)}{\partial x}=\frac{\partial
V(0;t,x,v)}{\partial v}=\alpha_2(s;t,x,v), \label{derivativexv}
\end{equation}
thus giving
\begin{equation}
\frac{\partial X(0;t,x,v)}{\partial x}=\frac{\partial
V(0;t,x,v)}{\partial v}\geq  \alpha_0>0 \label{derivativev}
\end{equation}
because of  assumption (\ref{force05}). Put $L(R)=L_0(R)\int_{{\bf
R}^3}m(v)dv/\alpha_0.$ Then, by use of  (\ref{derivativev}),
(\ref{factorf01}) gives (\ref{factorf}). This means that
(\ref{factorf01}) is a stronger assumption than (\ref{factorf}) when
the external forces satisfy the assumptions
(\ref{force01})-(\ref{force05}). Therefore the locally Lipschitz
assumption (\ref{factorf}) is at least mathematically very useful in
a more general case.

Now let us recall the Enskog equation for our further understanding
the locally Lipschitz assumption (\ref{factorf}) of the factor
$F^\pm$ under the assumptions (\ref{force01})-(\ref{force05}) of the
external forces. The Enskog equation can be roughly divided into two
classes: the standard and the revised one.  In the standard Enskog
equation (see \cite{cc70}, \cite{e22}, \cite{p89}), $F^\pm$ are
defined by a geometrical factor $Y$ which is
 a contact-value pair correlation function
of the hard-sphere system at uniform equilibrium and depends on the
density $\rho(t,x),$ i.e., they are given by
$F^+=Y(\rho(t,x-a\omega/2))$ and $F^-=Y(\rho(t,x+a\omega/2)).$ For a
fairly rare uniform gas of one particle with mass $m,$ it can be
found in \cite{cc70} that the value of $Y$ is approximatively
expressed by $Y(\rho(t,x))=[1-11b\rho(t,x)/8]/[1-2b\rho(t,x)]$ where
$b=2\pi a^3/(3m).$ It can be easily shown that in this case the
factor $F^\pm=Y$ satisfies (\ref{factorf01}) and is locally
Lipschitz as defined in (\ref{factorf}) with the external forces of
the assumptions (\ref{force01})-(\ref{force05}) for $R\in (0, R_0],$
where $R_0$ is some suitably small positive constant. Generally, the
dependence of the function $Y$ on the local density $\rho(t,x)$ is
of the form
\begin{equation}
Y(\rho(t,x))=1+\sum\limits_{i=1}^{+\infty}b_i[2\pi
a^3\rho(t,x)/3]^i, \label{ses}
\end{equation}
where $b_i$ are given in terms of the virial coefficients $B_i$
appearing in the equation of state for the hard sphere system. We
cannot know whether the series (\ref{ses}) converges when one of the
two different locally Lipschitz conditions (\ref{factorf}) and
(\ref{factorf01}) is satisfied. Even if this series converges, we
cannot yet know whether one of
 the assumptions (\ref{factorf}) and (\ref{factorf01}) of $F^\pm$ holds
 for any factor $Y$ of the above form.
Of course, if $F^\pm=Y$ is a factor defined by the form (\ref{ses})
with finite terms, then (\ref{factorf01}) holds and so $F^\pm$
satisfy (\ref{factorf}) when the external forces are of the
assumptions (\ref{force01})-(\ref{force05}).

In the case of the revised Enskog equation (see \cite{a90},
\cite{p06}, \cite{ve73}), $F^\pm$ are expressed by a contact-value
pair correlation function $G$ of the hard-sphere system at
non-uniform equilibrium. The form of $G$ is given by the Mayer
cluster expansion in terms of local density $\rho(t,x).$ The
function $G$ depends on the position $x,$  the vector $x\pm a\omega$
and the density $\rho(t,x),$ i.e., $F^+=G(x,x-a\omega, \rho(t,x))$
and $F^-=G(x,x+a\omega, \rho(t,x)).$
 In fact,  $G(x,y,\rho(t,x))=\exp(-\beta\Phi(|x-y|))\widetilde{G}(\rho(t,x)),$
where $\beta$ is a positive constant, $\Phi(|x-y|)$ is a potential
of two interaction spheres at the positions $x$ and $y,$ and
$\widetilde{G}$ is a functional of the following form \cite{p06}
\begin{eqnarray}
\widetilde{G}(\rho(t,x))=
1+\int V(12|3)\rho(3)dx_3+\frac{1}{2}\iint V(12|34)\rho(3)\rho(4)dx_3dx_4 \hspace*{0.17\linewidth}\nonumber\\
+\cdots+\frac{1}{(k-2)!}\iint\cdots\int V(12|34\cdots
k)\rho(3)\rho(4)\cdots \rho(k) dx_3dx_4\cdots dx_k+\cdots,
\label{ser}
\end{eqnarray}
here $\rho(k)=\rho(t,x_k) $ and $V(12|34\cdots k)$ is the sum of all
the graphs of $k$ labeled points which are biconnected when the
Mayer factor $\exp(-\beta\Phi(|x_i-x_j|))-1$ are added. In contrast
to the standard Enskog equation, the revised Enskog equation
possesses an H-function \cite{r78}. It can be also known that
 the revised Enskog theory for mixtures is consistent with irreversible thermodynamics,
including Onsager reciprocity relations (see \cite{ve73m},
\cite{v83},
 \cite{v90}). But we cannot know whether the above
series (\ref{ser}) converges under the assumption (\ref{factorf}) or
(\ref{factorf01}). We cannot yet know whether one of the assumptions
(\ref{factorf}) and (\ref{factorf01}) of $F^\pm$ holds for any
functional $\widetilde{G}$ of the above form.

The assumption (\ref{factorf}) or (\ref{factorf01}) of $F^\pm$ is
satisfied by some geometric factors present in the truncated Enskog
equations in both standard and revised cases. In the standard case
the geometric factor $Y$ considered above can be assumed to be of
the truncated form (\ref{ses}) with finite terms while in the
revised case an obvious example is that the functional
$\widetilde{G}$ can be truncated to be a positive constant. It can
be found that these factors satisfy the two assumptions of $F^\pm$
when one assumes that both $L(R)$ and $L_0(R)$ are two positive
constant functions on $R_+.$ Therefore the above two assumptions are
completely suitable for our understanding evolutions of moderately
or highly dense gases by use of our investigation of the properties
of the Enskog equation.

\bigskip

%\vspace*{\baselineskip}
{\small\noindent{\sc\bf Acknowledgement.}~The author would like to
thank the referees of this paper for their valuable comments and
suggestions on this work.}

%{\small

%}

\begin{thebibliography}{99}
\bibitem[1]{a90}Arkeryd L., On the Enskog equation with large initial data,
SIAM Journal on Mathematical Analysis, 1990, {\bf 21}: 631-646.
\bibitem[2]{ac89}Arkeryd L.,
Cercignani C., On the Convergence of Solutions of the Enskog
Equation to Solutions of the Boltzmann Equation, {Comm.}  in
 PDE, 1989, {\bf 14}: 1071-1089.
\bibitem[3]{ac90}Arkeryd L., Cercignani C., Global Existence  in  $L^1$  for  the
Enskog Equation and Convergence of the Solutions to Solutions  of
the Boltzmann Equation, {J.} {Stat.} {Phys.}, 1990, {\bf 59}:
845-867.
\bibitem[4]{bt85}Bellomo N., Toscani G., On the Cauchy problem for the nonlinear
Boltzmann equation. Global existence, uniqueness and asymptotic
stability, J.~Math.~Phys., 1985, {\bf 26}: 334-338.
\bibitem[5]{c87}Cercignani C., Existence of Global Solutions  for the Space
Inhomogeneous Enskog Equation, {Transp.} Theory {Stat.}  {Phys.},
1987, {\bf 16}: 213-221.
\bibitem[6]{c88}Cercignani C., Small Data Existence for the Enskog
equation  in $L^1$, {J.} {Stat.} {Phys.}, 1988, {\bf 51}: 291-297.
\bibitem[7]{cc70}Chapman S., Cowling T.~G., The Mathematical Theory of Non-Uniform Gases, Cambridge University Press, Third Edition, 1970.
\bibitem[8]{dl}DiPerna R.~J., Lions P.~L., On the Cauchy problem for  Boltzmann
equations: Global existence and weak stability, Ann.~Math., 1989,
{\bf 130}: 321-366.
\bibitem[9]{dyz05}Duan R., Yang T., Zhu C., Global existence to
Boltzmann equation with external force in infinite vacuum, Journal
of Mathematical Physics, 2005, {\bf 46}, 053307.
\bibitem[10]{e22}Enskog D., Kinetiche Theorie der W\`{a}rmeleitung,
Reibung und Selbstdiffusion in gewissen werdichteten Gasen und
Fl$\ddot{u}\beta$igkeiten, Kungl. Sv. Vetenskapsakademiens Handl. 63
(1922), 3-44, English Transl. in Brush, S.~G., Kinetic Theory, vol
3, Pergamon, New York 1972.
\bibitem[11]{guo01}Guo Y., The Vlasov-Poisson-Boltzmann system near vacuum,
Comm.~Math.~Phys., 2001, {\bf 218}: 293-313.
\bibitem[12]{is84}Illner R., Shinbrot M., The Boltzmann Equation, global existence
for a rare gas in an infinite vacuum, Comm.~Math.~Phys., 1984, {\bf
95}: 217-226.
\bibitem[13]{j07}Jiang Z., Global Solution to the Relativistic Enskog
Equation with Near-Vacuum Data, Journal of Statistical Physics,
2007, {\bf 127}: 805-812.
\bibitem[14]{ks78}Kaniel S., Shinbrot M., The Boltzmann Equation I. Uniqueness and local existence,
Comm.~Math.~Phys., 1978, {\bf 58}: 65-84.
\bibitem[15]{p89}Polewczak J., Global existence and asymptotic behavior for the nonlinear Enskog equation,
SIAM Journal on Applied Mathematics, 1989, {\bf 49}: 952-959.
\bibitem[16]{p06}Polewczak J., On some open problems in the revised Enskog equation for dense gases,
in Proceedings ``WASCOM 99'' 10th Conference on Waves and Stability
in Continuous Media, Vulcano (Eolie Islands), Italy, 7-12 June 1999,
V. Ciancio, A. Donato, F. Oliveri, and S. Rionero, Eds., World
Scientific Publishing, London, 2001, 382-396.
\bibitem[17]{r78}Resibois P., H-Theorem for the (Modified) Nonlinear Enskog Equation, J. Stat. Phys., 1978, {\bf 19}: 593-609.
\bibitem[18]{t86}Toscani G., On the non-linear Boltzmann equation in unbounded domains,
Arch.~Rational Mech.~Anal., 1986, {\bf 95}: 37-49.
\bibitem[19]{ukai86}Ukai S., Solutions of the Boltzmann Equation, Studies in Math.~Appl., 1986, {\bf 18}: 37-96.
\bibitem[20]{ve73}van Beijeren H., Ernst M.~H., The modified Enskog equation,
Physica, 1973, {\bf 68}: 437-456.
\bibitem[21]{ve73m}van Beijeren H., Ernst M.~H., The modified Enskog equation for mixtures,
Physica, 1973, {\bf 70}: 225-242.
\bibitem[22]{v83}van Beijeren H., Equilibrium distribution of hard-sphere systems and revised Enskog theory, {Phys.} {Rev.} {Lett.},
 1983, {\bf 70}: 1503-1505.
\bibitem[23]{v90}van Beijeren H., Kinetic theory of dense gases and liquids, in {\it Fundamental Problems in Statistical Mechanics VII},
H.~van Beijeren {Ed.}, Elsevier, 1990, 357-380.
\end{thebibliography}
\end{document}